\documentclass[
               prd,nofootinbib,superscriptaddress,
               showpacs,preprintnumbers,
               amsmath,amssymb]
              {revtex4}
\usepackage{graphicx,color}
\usepackage{epsfig}

\def\be{\begin{equation}}
\def\ee{\end{equation}}
\newcommand{\bea}{\begin{eqnarray}}
\newcommand{\eea}{\end{eqnarray}}

\newcommand{\dm}{\partial_\mu}
\newcommand{\dn}{\partial_\nu}

\providecommand{\href}[2]{#2}

\begin{document}

\preprint{INR/TH-39-2004}
\title{Lorentz-violating graviton masses: getting around ghosts,
low strong coupling scale and
VDVZ discontinuity}


\author{V.~Rubakov}
\affiliation{Institute for
Nuclear Research of the Russian Academy of Sciences,\\
  60th October Anniversary prospect 7a, Moscow 117312, Russia}

\begin{abstract}

A theory with the action combining
the Einstein--Hilbert term and graviton mass
terms violating Lorentz invariance is considered at linearized level
about Minkowskian background. It is  shown that with 
one of the masses set equal to zero, the theory
has the following properties: (i) there is a gap of order $m$
in the spectrum, where $m$ is the graviton mass scale;
(ii) the dispersion relations at ${\bf p}^2 \gg m^2$
are $\omega^2 \propto {\bf p}^2$, the spectrum of tensor modes  being
relativistic, while other modes having unconventional maximum velocity;
(iii) the VDVZ discontinuity  is absent; (iv)
the strong coupling scale is $(mM_{Pl})^{1/2}$.
The latter two properties are in sharp contrast to the
Lorentz-invariant
gravity with the Pauli--Fierz mass term.

\end{abstract}

\maketitle

\section{Introduction and Summary}

Massive, Lorentz-invariant gravity about flat background
has a number of peculiar properties. 
For general graviton mass terms, the theory possesses
ghosts --- fields with  wrong sign of the kinetic term.
If the graviton mass terms are of the Fierz--Pauli form~\cite{FP},
the ghosts are absent, but in the limit of vanishing  graviton mass $m$,
the graviton propagator exhibits
the van~Dam--Veltman--Zakharov (VDVZ) discontinuity~\cite{VDV,Z} 
originating from
a scalar
degree of freedom which does not decouple in the massless limit. 
At the classical level this scalar may not be a problem due to
non-linear 
effects~\cite{Vainshtein:1972sx,Deffayet:2001uk,Gruzinov:2001hp}.
However, at the quantum level the theory becomes strongly 
coupled~\cite{AHGS}
at energy scale $(m^4 M_{Pl})^{1/5}$, which has been 
confirmed by explicit calculations~\cite{Aubert:2003je}.
By adding higher order operators, this scale can be
raised, but in any case it is
at best of order $(m^2 M_{Pl})^{1/3}$, which is well
below the naive expectation $(mM_{Pl})^{1/2}$. This story repeats
itself in brane-world models with gravity modified in the
infrared~\cite{Charmousis:1999rg,Kogan:1999wc,Dvali:2000hr,Dvali:2001ae}:
either there are 
ghosts~\cite{Pilo:2000et,Dubovsky:2002jm,Dubovsky:2003pn,Chacko:2003yp}
or  strong coupling occurs at 
energies well below a naive 
estimate~\cite{Luty:2003vm,Rubakov:2003zb} 
(see, however,
Refs.~\cite{Kolanovic:2003am,Porrati:2004yi,Nicolis:2004qq}).

In Minkowskian background, most of the models considered so far
exhibit Lorentz invariance. It has been pointed out
recently~\cite{Arkani-Hamed:2003uy}, however, that the Higgs mechanism
for gravity would most
likely involve the violation of Lorentz invarance,
and, indeed, a model has been constructed~\cite{Arkani-Hamed:2003uy} 
in which excitations about flat space-time have Lorentz-violating 
spectrum. This is not yet a model with massive graviton, since in the 
large range of spatial momenta ${\bf p}$,
the dispersion law is $\omega \propto {\bf p}^2$, while at very low
momenta, the frequency $\omega ({\bf p})$ 
is imaginary and the excitations grow. Still the
question arises whether the violation of Lorentz invariance may help
to obtain a theory of massive gravitons which does not have ghosts
and
VDVZ discontinuity, and whose strong coupling scale is $(mM_{Pl})^{1/2}$.
  
In this paper we adopt bottom-up approach, and simply consider
a deformation of GR by Lorentz-violating graviton mass terms,
about flat space-time.
We show that with one of the masses set equal to zero, the theory
possesses desirable properties (provided that some combinations
of other masses obey positivity conditions): there is a gap of order $m$
in the spectrum, 
the dispersion relations at ${\bf p}^2 \gg m^2$
are $\omega^2 \propto {\bf p}^2$ (the spectrum of tensor modes  is
relativistic, while other modes have unconventional maximum velocity),
the VDVZ discontinuity  is absent and the strong coupling scale is
 $(mM_{Pl})^{1/2}$.

Technically, both VDVZ discontinuity and low strong coupling scale
occur due to normalisation factors relating the original fields
to canonically normalised ones~\cite{AHGS}: some of these factors
are of order $(m^2 M_{Pl})^{-1}$ while a naive expectation would be
$(mM_{Pl})^{-1}$. Thus, after introducing notations in Section~2,
we consider linear
excitations in Section~3, emphasising the 
normalisation  issue. We find that the normalisation factors are
indeed at most
of order $(mM_{Pl})^{-1}$ for all kinds of modes, which
immdiately tells that the strong coupling scale is   $(mM_{Pl})^{1/2}$.
We also study the spectrum and find that it has a gap, again
for all kinds of modes. In Section~4 we analyse interaction between
conserved sources, at linearised level, 
and show explicitly that in the massless limit
it reduces to the GR form, i.e., there is no VDVZ
discontinuity.

Thus, GR with Lorentz-violating graviton masses is a healthy
theory. This suggests that there may exist a Higgs phase of gravity
which has Minkowskian background, violates Lorentz invarance,
describes massive (and/or unstable)
gravitons and has intrinsic energy
scale  $(mM_{Pl})^{-1}$.

\section{Lorentz-violating mass terms}
Let us consider the action
\be
   S = S_{EH} + S_{m}
\label{genac}
\ee
where $S_{EH}$ is the Einstein--Hilbert term and
$S_{m}$ is the graviton mass term that explicitly violates
the Lorentz symmetry, but does not violate the Euclidean symmetry
of the three-dimensional space. The corresponding Lagrangian is
\be
L_{m} = \frac{M_{Pl}^2}{2} [m_0^2 h_{00} h_{00}
+ 2m_1^2 h_{0i}h_{0i} - m_2^2 h_{ij} h_{ij}
+ m_3^2 h_{ii} h_{jj}
- 2 m_4^2 h_{00} h_{ii}]
\label{Mh}
\ee
Here $h_{\mu \nu}$ are perturbations about Minkowski metric.
In what follows we will make a comparison to the Fierz--Pauli case,
the corresponding Lagrangian being
\[
 L_{FP} = \frac{M_{Pl}^2}{2} [-m^2 h_{\mu \nu} h^{\mu \nu}
+ m^2 (h_\mu^\mu)^2]
\]
Thus, the Fierz--Pauli Lagrangian is recovered when all masses
in eq.~(\ref{Mh}), except for $m_0$, are equal,
\[
  \mbox{FP}: \;\;\; m_0^2 = 0 \;, \;\;
  m_1^2 =m_2^2 = m_3^2 =m_4^2 = m^2 
\]
The latter property explains the conventions used in eq.~(\ref{Mh}).

Throughout this paper we  study the case
\be
           m_0 = 0
\label{m0=0}
\ee
In this case the field $h_{00}$ enters the action linearly,
i.e., it acts as the Lagrange multiplier. We will see that
this property (plus inequalities involving other masses)
is sufficient to ensure that the theory is
free of ghosts. We will also assume that all other masses
are proportional to a single
scale  which we  generically denote by $m$.

The properties of perturbations  depend on the representation
of the Euclidean symmetry group. It is thus convenient to express 
$h_{\mu \nu}$ in terms of
irreducible fields,
\bea
    h_{00} &=& \psi 
\nonumber
\\
 h_{0i}  &=& u_i + \partial_i v
\nonumber
\\
 h_{ij} &=& \chi_{ij} + (\partial_i s_j + \partial_j s_i)
 + \partial_i \partial_j \sigma + \delta_{ij} \tau
\nonumber
\eea
Here $\chi_{ij}$ is tranverse-traceless (tensor modes); 
$u_i$ and $s_i$ are transverse (vectors), while other fields are
three-dimensional scalars. Under the 
gauge transformations of GR,
$h_{\mu\nu} \to h_{\mu \nu} + \dm \xi_\nu +\dn \xi_{\mu}$, 
the tensor modes are invariant, while vectors and scalars
transform in the following way,
\bea
    u_i && \to u_i + \partial_0 \xi^T_i
\nonumber
\\
    s_i && \to s_i + \xi_i^T
\nonumber
\eea
where  $\xi_i^T$ is the transvese part of $\xi_i$,
and
\bea
   \psi && \to \psi + 2 \partial_0 \xi_0
\nonumber
\\
   v && \to v+ \partial_0 \eta + \xi_0
\nonumber
\\
   \sigma && \to \sigma + 2\eta
\nonumber
\\
   \tau && \to \tau
\nonumber
\eea
where $\eta$ is the longitudinal part of $\xi$, i.e., $\xi_i^L =
\partial_i \eta$. There is one gauge-invaiant combination in the
vector sector,
\be
 w_i  = u_i - \partial_0 s_i
\label{w}
\ee
and two gauge-invarants in the scalar sector, namely,
$\tau$ and
 \be
\Phi  = \psi - 2\partial_0 v + \partial^2_0 \sigma
\label{Sigma}
\ee
Up to total derivatives, the quadratic part of
the Einstein--Hilbert Lagrangian
is expressed in terms of these gauge-invariant combinations,
\bea
L_{EH} =&&\frac{M_{Pl}^2}{2} \left(-\chi_{ij} \Box \chi_{ij} 
- 2 w_i \Delta w_i \right.
\nonumber
\\
  &+& \left. 4 \Phi \Delta \tau 
  + 6 \tau \partial_0^2 \tau
  - 2 \tau \Delta \tau \right)
\label{Emodes}
\eea
where $\Delta$ is the three-dimensional Laplacian,
while the mass term is
\bea
L_{m} &=& \frac{M_{Pl}^2}{2}\left[2m_1^2(u_i u_i
+ \partial_i v \partial_i v) \right.
\nonumber \\
&-& m_2^2(\chi_{ij}\chi_{ij} + 2\partial_i s_j \partial_i s_j
+ \partial_i \partial_j \sigma \partial_i \partial_j \sigma 
+ 2 \tau \Delta \sigma + 3 \tau^2) 
\nonumber \\
&+& \left.
m_3^2 (\Delta \sigma + 3 \tau)^2 - 2m_4^2 \psi (\Delta \sigma + 3\tau)
\right]
\label{Lm}
\eea 
We are now ready to study the physical excitations.

\section{Physical excitations and their normalisation}

\subsection{Tensors}

The field equation for tensors is
\be
 \Box \chi_{ij} + m_2^2 \chi_{ij} = 0
\label{tens}
\ee
Thus, these modes have relaivistic spectrum with mass $m_t = m_2$.
From the action (\ref{Emodes}) one observes that 
the normalisation factor is
$M_{Pl}^{-1}$, which is usual for gravitons. This means that 
strong coupling for tensors at low energies is only due to
their interactions with other modes. We will not consider
tensors any longer in this paper.

\subsection{Vectors}

It is clear from eqs.~(\ref{Emodes}) and (\ref{w}) that the field
$u_i$ enters the action without time derivatives, i.e., it is
a non-dynamical field. We integrate it out by using the field
equation obtained by varying the action 
with respect to  $u_i$ itself,
\be
 \Delta (u_i - \partial_0 s_i) - m_1^2 u_i 
\equiv \Delta w_i - m_1^2 u_i =0
\label{v1}
\ee
We get
\[
   u_i = \frac{\Delta}{\Delta - m_1^2} \partial_0 s_i
\]
Plugging this expression back into the action, we obtain 
for vector part
\be
  L_{v} = M_{Pl}^2
\left[ m_1^2 \partial_0 s_i  \frac{\Delta}{\Delta - m_1^2} 
\partial_0 s_i + m_2^2 s_i \Delta s_i \right]
\label{Lv}
\ee
Note that in momentum space
\[
\frac{\Delta - m_1^2}{\Delta} = \frac{{\bf p}^2 + m_1^2}{{\bf p}^2}
\]
is positive for positive $m_1^2$, so the term with time derivatives
has correct sign (the term with spatial gradient also has correct sign).

To obtain the canonical action, we define a new field $s^c_i$ such that
\[
  s_i = \frac{1}{m_1 M_{Pl}} \sqrt{\frac{\Delta - m_1^2}{2 \Delta}} s^c_i
\]
and get
\[
 L_{v} = 
\frac{1}{2} \left[ \partial_0 s^c_i  
\partial_0 s^c_i -  \frac{m_2^2}{m_1^2} \partial_j s^c_i \partial_j s^c_i 
- m_2^2 s^c_i s^c_i\right]
\]
Thus, the spectum is
\[
   \omega^2 = \frac{m_2^2}{m_1^2} {\bf p}^2 + m_2^2
\]
In terms of the canonically normalised field $s^c_i$, the original fields
$s_i$ and $u_i$ are proportional to $(mM_{PL})^{-1}$. 
The only gauge-invariant 
combination in the vector sector, $w$, written in terms of 
the canonicaly normalised field is of order
\be
      w_i \propto \frac{m}{M_{Pl}} s^c_i 
\label{worder}
\ee  
This follows, e.g., from eq.~(\ref{v1}). This behavior of the vector
modes at small $m$ 
is precisely the same as in the Fierz--Pauli case;
in paticular,
the analysis of Ref.~\cite{AHGS} suggests 
that strong coupling occurs at
\be
    E \sim \sqrt{mM_{Pl}}
\label{strongcoupling}
\ee
which is a relatively high scale.

Thus, for tensor and vector fields 
nothing changes, as compared to the Fierz--Pauli case,
except for the ``speed of light'' for vectors. 
The only constraints we have up to now are
\[
   m_1^2 > 0\; , \;\;\;\; m_2^2 > 0
\]
Then both vectors and tensors are massive positive energy fields.

\subsection{Scalars}

As we already pointed out, we will consider ghost-free case
(\ref{m0=0}). In this case
the field $\psi \equiv h_{00}$ 
is the Lagrange multiplier. The corresponding constraint is
\be
   \frac{\delta S}{\delta \psi} \propto
2 \Delta \tau + m_0^2 \psi - m_4^2 (\Delta \sigma + 3\tau) = 0
\label{s1}
\ee
We use this constraint to express $\sigma$
through $\tau$,
\be
  \sigma = \frac{2}{m_4^2} \tau - \frac{3}{\Delta} \tau
\label{sigma}
\ee
Now, the field $v$ enters the action without time derivatives
(the term proptional to $\partial_0 v \Delta \tau$ in (\ref{Emodes})
may be written as $v \partial_0 \Delta \tau$  ). Thus, we eliminate 
this field by making use of its field equation,
\be
 \frac{\delta S}{\delta v} \propto
    2\partial_0 \tau - m_1^2 v= 0
\label{s2}
\ee
and find
\be
       v=\frac{2}{m_1^2} \partial_0 \tau
\label{v}
\ee
We plug the expressions (\ref{sigma}) and (\ref{v}) back
into the action, and obtain the action in terms of the only 
remaining dynamical field $\tau$. The corresponding Lagrangian is
\bea
L_\tau &=& \frac{M_{Pl}^2}{2}\left[\left(\frac{8}{m_4^2} - \frac{8}{m_1^2}
\right)\Delta \tau \partial_0^2 \tau
 - 4\frac{m_2^2 - m_3^2}{m_4^4} (\Delta \tau)^2 \right.
\nonumber \\
&-& \left. 6 \tau \partial_0^2 \tau + \left(8 \frac{m_2^2}{m_4^2} -2\right) 
\tau \Delta \tau - 6m_2^2 \tau^2 \right]
\label{remarkable}
\eea
This is the central formula of this section. 
It enables one to immediately derive both the spectrum and
normalisation factor 
relating $\tau$ and canonically normalised field $\tau^c$,
\[
   \tau = \frac{1}{M_{Pl}}\cdot
\left[-\left(\frac{8}{m_4^2} - \frac{8}{m_1^2}
\right)\Delta + 6\right]^{-\frac{1}{2}} \cdot \tau^c 
\propto \frac{m}{M_{Pl}} \tau^c
\]
In 
the general case the Lagrangian (\ref{remarkable})
contains terms
enhanced by $1/m^2$, which explains why  the normalisation factor for
$\tau$ is proportional to $m/M_{Pl}$. 
The largest fields $\sigma$ and $v$ are propotional to 
$1/(M_{Pl}m)$,
in a complete analogy to vector modes.  To understand the
properties of another gauge-invariant field $\Phi$,
we make use of a linear combination of the two remaining field
equations, 
\[
 \Phi -\tau + m_2^2 \sigma = 0
\]
Fom eq.~(\ref{sigma}) we deduce that
\[
\Phi \propto \frac{m}{M_{Pl}} \tau^c
\]
Thus, both gauge-invariant 
fields $\tau$ and $\Phi$, 
expressed through the canonically normalised field $\tau^c$,
are of order $\frac{m}{M_{Pl}}$,
again in 
analogy to the vector case, eq.~(\ref{worder}). All this ensures 
that the strong coupling scale in the scalar sector is the
same as in the vector sector, and is given by eq.~(\ref{strongcoupling}), 
and suggests that there is no VDVZ discontinuity.

For
\be
      m_1^2 > m_4^2 > 0 \; , \;\;\; m_2^2 > m_3^2
\label{general}
\ee
and
\[
     4m_2^2 > m_4^2
\]
all terms in the action have correct signs (recall that $\Delta$ is
negative-definite). Thus, there are no ghosts or tachyons.
The spectrum is 
\[
     \omega^2 = \frac{{\bf p}^2 + z \mu_1^2}{{\bf p}^2 + \mu_0^2}\cdot
 \frac{\mu_0^2}{\mu_1^2}\cdot {\bf p}^2 + 
\frac{m_2^2 \mu_0^2}{{\bf p}^2 + \mu_0^2}
\]
where the parameters are all positive and are defined as follows,
\[
   \frac{4}{m_4^2} - \frac{4}{m_1^2} = \frac{3}{\mu_0^2}\; ,\;\;\;
    2 \frac{m_2^2 - m_3^2}{m_4^4} = \frac{3}{\mu_1^2} \; , \;\;\;
  4\frac{m_2^2}{m_4^2} -1 = 3z
\] 
At high momenta one has
\[
\omega^2 =   \frac{\mu_0^2}{\mu_1^2}\cdot {\bf p}^2 \; , \;\;\;
{\bf p}^2 \gg m^2
\]       
while at low momenta there is a gap
\[
   \omega^2 = m_2^2 \; , \;\;\;\; {\bf p} = 0
\]
In this sense the scalar mode is also massive.

Let us now compare our results to the Lorentz-invariant case.
In that case, the relation $m_0=0$ implies the Fierz--Pauli
form of the mass terms.
In the Fierz--Pauli case the terms in the first line in 
eq.~(\ref{remarkable}) vanish, so $\tau \propto 1/M_{Pl}$
in terms of canonically normalised field, and 
$v, \sigma \propto 1/(M_{Pl}m^2)$. This is in agreement
with Ref.~\cite{AHGS}, and imples the VDVZ discontinuity, 
as well as
the low energy scale of
strong coupling. Needless to say, the action for $\tau$ takes
the Lorentz-invariant form.

\subsection{High-energy limit}

To end up this section, let us mention an alternative
way of obtaining the above 
properties of the three-vector and scalar
modes in the high-energy regime $E \gg m$.
To analyse the quadratic action directly in the high-energy
limit, one makes use of the formalism of the St\"uckelberg type.
The relevant part of the metric perturbations is  
``pure gauge''~\cite{AHGS},
\be
    h_{\mu \nu} = \partial_\mu \pi_\nu + \partial_\nu \pi_\mu
\label{St}
\ee
and the relevant part of the Lagrangian is the mass term (\ref{Mh}).
For the field of the form (\ref{St}) the Lagrangian 
is, after integrating by parts
 (we still set $m_0=0$),
\be
   L_{m} = \frac{M_{Pl}^2}{2}
[ 2m_1^2 (\partial_0 \pi_i)^2 + 2m_1^2 (\partial_i \pi_0)^2
+ (8 m_4^2 - 4 m_1^2) \pi_0 \partial_0 \partial_i \pi_i
- 2m_2^2 (\partial_i \pi_j)^2
- (2m_2^2 - 4m_3^2) (\partial_i \pi_i)^2]
\label{StL}
\ee
For the transverse part, $\pi_i^T = s_i$, one immediately 
obtains that this Lagrangian coincides with the high-energy 
limit of the Lagrangian (\ref{Lv}). In the scalar part, the field  
$\pi_0$ is non-dynamical, and may be eliminated by making use of 
its field equation,
\[
  \pi_0 = \frac{2m_4^2 - m_1^2}{m_1^2 \Delta} 
\partial_0 \partial_i \pi_i
\]
Then the longitudinal part of the Lagrangian (\ref{StL})
becomes
\[
 L^{L}= \frac{M_{Pl}^2}{2}
\left[8 m_4^4 \left(\frac{1}{m_4^2} -\frac{1}{m_1^2} \right)
(\partial_0 \pi_i^L)^2
- 4 (m_2^2 - m_3^2)(\partial_i \pi_i^L)^2 \right]
\]
This coincides with the Lagrangian (\ref{remarkable}), if one
identifies
\[
   \pi_i^L = \frac{1}{2} \partial_i \sigma
\]
and recalls eq.~(\ref{sigma}) which in the high-energy limit 
reduces to $\sigma = 2\tau/m_4^2$.

Thus, the Lorentz-violating mass terms give rise 
to healthy kinetic terms for all components of $\pi$'s.
Repeating the analysis of Ref.~\cite{AHGS}, it is straightforward
to see that the strong-coupling scale is indeed the same in the
transvese and longitudinal sectors, and is given by 
eq.~(\ref{strongcoupling}). This is in sharp contrast to the 
Fierz--Pauli case, in which the mass terms per se do not
produce the kinetic term for the longitudinal part of 
$\pi$~\cite{AHGS}.

Finally, let us briefly discuss the case $m_0 \neq 0$.
In that case, the term $4m_0^2 (\partial_0 \pi_0)^2$ is added
to the Lagrangian (\ref{StL}), so the field $\pi_0$ becomes
dynamical. Depending on the sign of $m_1^2$, either the 
kinetic term for $\pi_i$ (the first term in eq.~(\ref{StL}))
or the gradient term for $\pi_0$ 
(the second term in eq.~(\ref{StL})) has wrong sign, so the energy
is unbounded from below. The case $m_0 \neq 0$, $m_1 = 0$ may 
be interesting. In that case the fields $\pi_i$ are non-dynamical;
after integrating them out at the level of the effective Lagrangian
(\ref{StL}), no terms with spatial gradient of $\pi_0$
appear. By analysing the complete linearised theory, one finds that
there are no physical excitations\footnote{Even more special case is 
$m_0 \neq 0$, $m_1 =0$ and $m_4^2 = m_2^2 - m_3^2$, in which the
quadratic action possesses an accidental (?) gauge symmetry.}   
in the scalar and vector sectors,
yet tensor modes obey eq.~(\ref{tens}) and have physical
exciations of mass $m_t = m_2$. We think that
the theory with $m_0 \neq 0$ and
$m_1 = 0$ deserves further study.

\section{Interaction with conserved sources}

To see explicitly that the VDVZ discontinuity 
is absent in the interesting case
(\ref{general}), let us study interactions between
conserved sources. In GR, the field
induced by conserved energy-momenum $T_{\mu\nu}$
is
\[
   \mbox{GR:} \;\;\;  h_{\mu \nu} = -\frac{1}{\Box}
\left(t_{\mu \nu} - \frac{1}{2} t^\lambda_\lambda\right) + \dots
\]
where dots denote  
longitudinal terms irrelevant for the interaction between
conserved sources, and
\[
t_{\mu \nu} = \frac{1}{M_{Pl}^2}T_{\mu \nu}
\]
On the other hand,
in the Fierz--Pauli case in the limit $m \to 0$, one has
\[
\mbox{FP:} \;\;\;     h_{\mu \nu} = -\frac{1}{\Box}
\left(t_{\mu \nu} - \frac{1}{3} t^\lambda_\lambda\right) +\dots
\]
The difference between the two expessions 
is precisely the VDVZ discontinuity. 

In terms of the gauge-invariant three-vector and scalar
fields, the corresponding 
expressions are
\bea
\mbox{GR~and~FP:} \;\;\;
  w_i &=& 
\frac{1}{\Delta} t_{0i}
\label{wE}
\\
 \mbox{GR:} \;\;\;  \tau&=& \frac{1}{2\Delta} t_{00}
\label{tauE}
\\
 \Phi &=&  \frac{1}{2\Delta} \left(t_{ii}+t_{00} - \frac{3}{\Delta}
\partial_0^2 t_{00} \right) 
\label{SigmaE}
\\
\mbox{FP:} \;\;\; 
\tau &=& \frac{1}{2\Delta}t_{00} + \frac{1}{6 \Box}(t_{00} - t_{ii})
\label{tauFP}
 \\
\Phi &=& \frac{1}{2\Delta} \left(
t_{ii} + t_{00} - \frac{3}{\Delta}\partial_0^2 t_{00}
\right) - \frac{1}{6 \Box}(t_{00} - t_{ii})
\label{SigmaFP}
\eea
Let us see that the masless limit of the theory with Lorentz-violaing 
mass terms
reproduces the GR expressions (\ref{tauE}) and (\ref{SigmaE}), as 
well as eq.~(\ref{wE}) for the vector part. By massless limit
we mean 
\[
 m_i^2 \to 0 \; , \;\;\;\; \frac{m_i^2}{m_j^2} = \mbox{fixed}
\; , \;\;\; i,j = 1, \dots, 4
\]
We still consider the
ghost-free case (\ref{m0=0}).

The reason for emphasising the gauge-invariant fields is as follows.
At the  linearised level the source term is
\[
S_{int} = - \int~d^4x~T_{\mu \nu} h^{\mu \nu}
\]
Making use of the conservation equations, $\partial_\mu T^\mu_\nu = 0$,
one expresses this action through gauge-invariant variables (\ref{w})
and (\ref{Sigma}),
\be
S_{int} = - \int~d^4 x~ (T_{ij}\chi_{ij} - 2T_{0i}w_i + T_{00} \Phi
+ T_{ii} \tau)
\label{Sint}
\ee
The interaction between  conserved sources
is thus determined by the gauge-invariant fields produced. 
The tensor part is trivial, so our purpose is
to calculate the fields $w_i$, $\Phi$ and $\tau$ generated by
the conserved source $T_{\mu \nu}$.

\subsection{Vectors}
Making use of eqs.~(\ref{Emodes}) and (\ref{Lm}), one varies 
the action (\ref{genac}), with the source term (\ref{Sint})
added, with respect to $u_i$ and $s_i$, and obtains the following
equations,
\be
 \Delta (u_i - \partial_0 s_i) - m_1^2 u_i = t_{0i}
\label{v1-t}
\ee
\be
 \partial_0 u_i - \partial_0^2 s_i - m_2^2 s_i = \frac{1}{\Delta} 
\partial_0 t_{0i}
\label{v2-t}
\ee
From eq.~(\ref{v1-t}) we get
\[
   u_i = \frac{\Delta}{\Delta - m_1^2} \partial_0 s_i + 
\frac{1}{\Delta - m_1^2} t_{0i}
\]
Plugging this into eq.~(\ref{v2-t}) we obtain
\[
 s_i =- 
\frac{1}{\Delta(\partial_0^2  -\frac{m_2^2}{m_1^2} \Delta + m_2^2)}
 \partial_0 t_{0i}
\]
which is finite in the massless limit.
The gauge-invariant combination is
\[
 w_i = \frac{m_1^2}{\Delta - m_1^2} \partial_0 s_i +
\frac{1}{\Delta - m_1^2} t_{0i}
\]
It has smooth massless limit,
which coincides with the GR expression (\ref{wE}).

\subsection{Scalars}

In the scalar sector, the field equations read
\bea
2 \Delta \tau - m_4^2 (\Delta \sigma + 3\tau) &=& t_{00}
\label{s1-t}
\\
    2\partial_0 \tau - m_1^2 v &=& \frac{1}{\Delta} \partial_0 t_{00}
\label{s2-t}
\\
  2 \partial_0^2 \tau - m_2^2 \Delta \sigma - m_2^2 \tau
  + m_3^2 \Delta \sigma + 3 m_3^2 \tau - m_4^2 \psi &=& \frac{1}{\Delta}
\partial_0^2 t_{00}
\label{s3-t}
\\
   2\Delta ( \psi -2\partial_0 v + \partial_0^2 \sigma) 
- 2 \Delta \tau + 2 m_2^2 \Delta \sigma &&
\nonumber
\\
\equiv 2 \Delta \Phi - 2\Delta \tau+ 2 m_2^2 \Delta \sigma &=& 
t_{ii} - \frac{3}{\Delta} \partial_0^2 t_{00}
\label{s4-t-short}
\eea
We note in passing that eq.~(\ref{s3-t}) is obtaned by varying
the action with respect to $\sigma$, while the variation with 
respect to $\tau$ gives a linear combination of eq.~(\ref{s4-t-short})
and eq.~(\ref{s3-t}), rather than eq.~(\ref{s4-t-short}) itself.

Equations (\ref{s1-t})~--~(\ref{s4-t-short}) are straightforwardly
solved.
Equation (\ref{s1-t}) gives
\be
   \sigma = \frac{2}{m_4^2} \tau - \frac{1}{m_4^2 \Delta} t_{00}
- \frac{3}{\Delta} \tau
\label{sigma-t}
\ee
From eq.~(\ref{s2-t}) we find
\be
       v=\frac{2}{m_1^2} \partial_0 \tau - \frac{1}{m_1^2\Delta}
\partial_0 t_{00} 
\label{tau-t}
\ee
while eq.~(\ref{s3-t}) yields
\[
\psi=\frac{1}{m_4^2}\left(2\partial_0^2 \tau - 
2 \frac{m_2^2 - m_3^2}{m_4^2} \Delta \tau + 2 m_2^2 \tau
+ \frac{m_2^2 - m_3^2}{m_4^2} t_{00} - \frac{1}{\Delta} \partial_0^2
t_{00} \right)
\]
Plugging these expressions into eq.~(\ref{s4-t-short})
we obtain
\bea
\left(\frac{8}{m_4^2} -\frac{8}{m_1^2} \right)
\Delta \partial_0^2 \tau
- 4 \frac{m_2^2 - m_3^2}{m_4^4} \Delta^2 \tau
 - 6 \partial_0^2 \tau +\left(8 \frac{m_2^2}{m_4^2} -2\right)
\Delta\tau
-6m_2^2 \tau &&
\nonumber \\
=  \left(\frac{4}{m_4^2} -\frac{4}{m_1^2} \right)
\partial_0^2 t_{00} -2 \frac{m_2^2 - m_3^2}{m_4^4}
\Delta t_{00}
 + t_{ii} - \frac{3}{\Delta} \partial_0^2 t_{00} 
+2 \frac{m_2^2}{m_4^2} t_{00} &&
\label{long}
\eea
Of course the left hand side of this equation matches the
Lagrangian (\ref{remarkable}). The point is that the coefficients in
the right hand side are such that  in the massless limit, the GR 
expression (\ref{tauE}) is reproduced (provided that $m_1^2 \neq
m_4^2$ and/or $m_2^2 \neq m_3^2$), 
i.e., $\tau = t_{00}/(2\Delta) + O(m^2)$.
Now, from the latter expession and eq.~(\ref{sigma-t}) it follows that
  $ \sigma$ is finite in the massless limit.
Then from eq.~(\ref{s4-t-short}) one finds that
$\Phi$ also has massless limit which 
coincides with the GR expression
(\ref{SigmaE}).  Thus, there is no VDVZ discontinuity.

As a cross check, one recovers the Fierz--Pauli case
by setting all masses in eq.~(\ref{long})
equal to each other. 
The resulting equation is straightforwardly solved; then 
using eqs.~(\ref{s4-t-short}) and
(\ref{sigma-t}), one indeed finds that the expressions
(\ref{tauFP}) and (\ref{SigmaFP}) are obtained in the massless
limit.

The author is indebted to S.~Dubovsky, D.~Krotov, M.~Porrati, 
R.~Rattazzi and S.~Sibiryakov for
encouraging correspondence.
He thanks the Abdus Salam International Centre for
Theoretical Physics, where part of the work has been done,
for hospitality. This work was supported 
in part by RFBR grant 02-02-17398.

\end{document}